\documentclass{article}
\usepackage{spconf,amsmath,graphicx,multirow,float,subcaption}
\usepackage{xcolor}
 
\title{An Investigation of Data Augmentation and One-Class Classifiers for Voice Replay Attack Detection}
\title{Audio compression-assisted feature extraction for voice replay attack detection}
\name{Xiangyu Shi$^{1*}$~\thanks{$^*$Equal Contribution}, Yuhao Luo$^{1*}$, Li Wang$^{1}$, Haorui He$^{1}$, Hao Li$^{2}$, Lei Wang$^{3}$, Zhizheng Wu$^{1}$}

\address{$^1$School of Data Science, Shenzhen Research Institute of Big Data, \\The Chinese University of Hong Kong, Shenzhen (CUHK-Shenzhen), China \\ $^2$Huawei Technology, China \quad $^3$Huawei International, Singapore}
\begin{document}
\ninept
\maketitle
\begin{abstract}
Replay attack is one of the most effective and simplest voice spoofing attacks. Detecting replay attacks is challenging, according to the Automatic Speaker Verification Spoofing and Countermeasures Challenge 2021 (ASVspoof 2021), because they involve a loudspeaker, a microphone, and acoustic conditions (e.g., background noise). One obstacle to detecting replay attacks is finding robust feature representations that reflect the channel noise information added to the replayed speech.
This study proposes a feature extraction approach that uses audio compression for assistance. Audio compression compresses audio to preserve content and speaker information for transmission. The missed information after decompression is expected to contain content- and speaker-independent information (e.g., channel noise added during the replay process).
We conducted a comprehensive experiment with 
3 classifiers on the ASVspoof 2021 physical access (PA) set and confirmed the effectiveness of the proposed feature extraction approach. \textbf{\textit{To the best of our knowledge, the proposed approach achieves the lowest EER at 22.71\% on the ASVspoof 2021 PA evaluation set.}}



\end{abstract}

\section{Introduction}
\vspace{-3mm}
\label{sec:intro}

\begin{figure*}[!h]
    \centering
    \includegraphics[width=\linewidth]{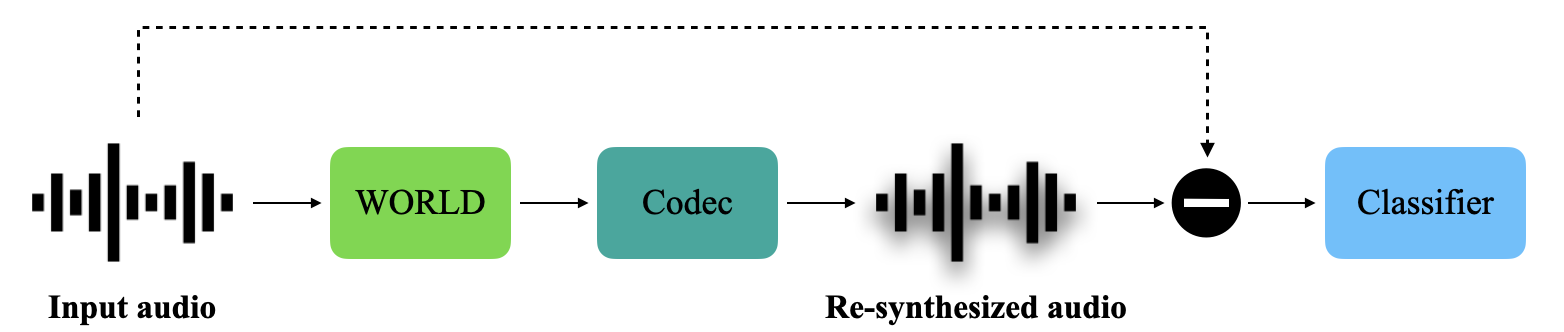}
    \caption{The overall framework of the replay detection system with assistance of audio compression (audio codec) for feature extraction. WORLD vocoder synthesizes the input audio and Opus codec will compress and then decompress the synthesized audio. We subtract the original audio and the re-synthesized audio in Mel-spectrogram form before classification}
    \label{fig:pipeline}
\end{figure*}



A replay attack is when a prerecorded speech sample is used to deceive a speaker verification system. This type of attack can be carried out with just recording and playback devices, and does not require expert knowledge. Studies show that the success rate of these attacks increases when the recording and playback devices produce more accurate reproductions of the original audio, making them sound more like a human voice.

There is increasing research interest in detecting replay attacks. The benchmark dataset for replay attacks was released during the ASVspoof competition, which took place in 2017~\cite{kinnunen2017asvspoof}, 2019~\cite{todisco2019asvspoof}, and 2021~\cite{yamagishi2021asvspoof}, spanning three editions of the competition. Currently, most research articles utilize the 2019 or 2021 dataset as the benchmark. It is worth noting that the 2021 edition provided only a test set, while the training set used the dataset from 2019.

Replay attack detection involves identifying channel noise that is produced during recording and playback. Current methods focus on extracting features and performing binary classification to determine whether the audio is genuine or spoofed. Researchers have found that audio features based on the Constant-Q Transform (CQT) are effective in detecting spoofing. Methods like CQCCE~\cite{yang2020significance} and CDOC~\cite{xu2023device}  have both been shown to be effective in this regard. Hidden artifacts in audio, such as silent segments and spectral defects, are crucial for detection~\cite{xue2023learning}. Channel noise from loudspeakers is used to learn patterns and detect spoofing in low-frequency signals~\cite{lu2022robust}. However, binary classification models face domain mismatch issues due to device diversity. In~\cite{wang2021dkucmri}, a generative model is used to only model bonafide samples for better discriminative ability. Additionally, a WORLD vocoder is used to remove speech components and improve the detection of spoofed audio.

According to ASVspoof 2021, the best system achieved 24.25\% in the challenge for the PA task. This suggests that existing feature representations are still not robust enough to detect replay attacks. Since a replay attack involves a loudspeaker, a microphone, and acoustic conditions such as background noise, the variations of devices (i.e. loudspeaker and microphone) and acoustic conditions can create unlimited versions of replayed speech. \textit{\textbf{One obstacle to detecting replay attacks is finding robust feature representations that reflect the channel noise information added to the replayed speech}}. 
The robust feature representation is expected to be independent of content and speaker information.

This study continues the quest for a robust feature representation and proposes a feature extraction approach that uses audio compression for assistance.
Audio compression compresses audio to preserve content and speaker information for transmission. \textit{\textbf{The information missed after decompression is expected to contain content- and speaker-independent information (e.g., channel noise added during the replay process)}}. 
With the assistance of audio compression, we also explore the effect of different kinds of data augmentation techniques and compare their performance
with three different one-class classifiers, namely variational autoencoder (VAE), one-class support vector machine (SVM), and AnoGAN -- a deep convolutional generative adversarial network.

\section{Feature extraction with the assistance of audio compression}
\vspace{-3mm}
\label{sec:augmentation}

In this study, we propose a feature extraction approach that uses audio compression for assistance. The overall framework with the proposed feature extraction approach is presented in Fig.~\ref{fig:pipeline}. We will describe the proposed framework, data augmentation methods and classifiers in this section. Data augmentation has improved the performance of many speech-related tasks. We examine the effectiveness of data augmentation along with the proposed framework.


\subsection{Audio Codec/Compression}

In this study, we investigate the use of audio codecs
for feature extraction. Audio codecs have traditionally been used for audio compression and decompression. Our work uses a similar approach to that of~\cite{dku}, 
which uses a WORLD vocoder~\cite{morise2016world} to assist with feature extraction by computing the differences between the vocoded and original waveforms as feature presentations. We argue that \textit{incorporating an audio codec could better preserve the most essential speech content information, with the difference between the original waveform and the decompressed audio reflecting content- and speaker-independent information more accurately}.
\textbf{This property makes it more suitable for replay detection}. 

To achieve our goal, we use the Opus Codec\footnote{https://opus-codec.org/} in this work. The Opus Codec is well-known for its adaptability and is widely used in real-time communication applications. For feature extraction,
we utilize the Opus encoder to compress the audio into packets at a specific bitrate. Then, we use the Opus decoder to unpack these packets and restore the audio to its original form. When the bitrate is lower than that of the original audio, it is expected that the Opus-processed audio will retain the complete vocal information while effectively reducing most of the channel noise.
As a result, we can distinguish between authentic and spoofed audio by comparing the original and the Opus-processed audio. We now compare Pyworld-Opus-processed audio instead of Opus-processed.

More specifically, we follow the following steps for feature extracting when using opus codec:
\begin{itemize}
    \item First, use the WORLD vocoder to resynthesize audio and use the Opus encoder to compress an original audio into a package.
    \item Next, employ the Opus decoder to decompress the package back into audio, resulting in an Opus-preprocessed audio file.
    \item After that, Transform both the original audio and the Opus-preprocessed audio into spectrograms.  
    \item Finally, subtract the temporal-averaged spectrograms from each other to obtain the feature representation.
\end{itemize}

\subsection{Data Augmentation}

Data augmentation has improved the performance of many speech-related tasks. In this work, we also study the impact of data augmentation with the proposed feature extraction.
The investigated data augmentation approaches are 
\begin{itemize}
    \item \textbf{SpecAugment}: SpecAugment~\cite{specaugment} is proposed for automatic speech recognition by applying masking to a spectrogram for data augmentation and robustness. In this study, we investigate \textbf{\textit{frequency masking}} and \textbf{\textit{time masking}}. Frequency masking is to apply masking to certain frequency bands, and time masking remove or reduce audio signals of specific time intervals to create gaps or silences in the waveform. 
    \item \textbf{AddNoise}: To add real office noise to audio. The noise used in this study contains the voices of multiple people talking. Noise is extracted from the VOiCES dataset~\cite{voices}.
    \item \textbf{AddReverb}: To simulate reverberation in a meeting room scenario. More specifically, Room Impulse Response (RIR) is convoluted with original audios. RIR data are from the VOiCES dataset~\cite{voices}.
    \item \textbf{AdjustSpeech}: We apply speed perturbation to randomly slow down or speed up audio by uncertain factors. Speed Perturbation factors are set to $[0.9, 1.1]$.
    \item \textbf{Pre-emphasis \& De-emphasis}: Manipulate audio to enhance the high-frequency components in audio. This is motivated by the findings presented in~\cite{lu2022robust}, which highlight that high-frequency components tend to contain more forged information, thereby benefiting replay attack detection.
\end{itemize}

\subsection{Classifier}

We also explore the performance of three one-class classifiers for detection with the proposed feature extraction. The classifiers are  variational auto-encoder (VAE), one-class support vector machine (SVM), and AnoGAN, a deep convolutional generative adversarial network. 

\subsubsection{Variational Auto-Encoder (VAE)}

The Variational Auto-encoder (VAE)~\cite{vae} can learn a low-dimensional representation of input data, often used in anomaly detection. The latent space, formed by the representation of input data, serves as a compressed representation that captures essential features of the input. The VAE is trained by minimizing a reconstruction loss, which measures dissimilarity between the original input and the reconstructed output. The reconstruction probability is used as the anomaly score. 

\textbf{Implementation details:} In this study, five linear layers are employed for encoder and decoder, respectively. ReLU is employed as an activation function. The size of the latent space is $2$, and the number of samples in the latent space is $10$.


\subsubsection{One-Class Support Vector Machine (OCSVM)}

The One-Class Support Vector Machine (OCSVM) was first introduced by Schölkopf et al.~\cite{OCSVM}. It allows SVM to be trained with only genuine data. OCSVM tries to separate all data points from the origin in a high-dimensional feature space using a hyperplane, as shown in the objective function below. Points lying below the hyperplane and closer to the origin are viewed as outliers.

\textbf{Implementation details:} We adopt OCSVM in scikit-learn\footnote{https://scikit-learn.org}, and keep the same default configuration(kernel=RBF, $\gamma$=scale, tol=1e-3, $\nu$=0.5). Note that $\nu$ is an upper bound on the fraction of training errors and a lower bound of the fraction of support vectors.

\subsubsection{AnoGAN: Anomaly GAN}

AnoGAN~\cite{AnoGAN} is a deep convolutional generative adversarial network, which was proposed in~\cite{AnoGAN} for anomaly detection. The AnoGAN uses two different loss functions: a weighted average of residual loss and a discrimination loss. The residual loss measures the difference between the real feature $\mathbf{x}$ and the generated feature $G(\mathbf{z})$, where $\mathbf{z}$ is a random input for generator.

\textbf{Implementation details:} The generator consists of 5 transposed convolutional layers. The kernel size, stride and padding are the same for all the layers, which are 4, 2 and 1, respectively. 2 linear layers are concatenated to the input layer and the output layer, respectively, to match the feature shape of generated data and real data. The discriminator consists of 5 convolutional layers. The kernel size, stride and padding are the same for all the convolutional layers, which are 4, 2 and 1, respectively. The weight coefficients of residual loss and discrimination loss are equal, both are 0.5.

\begin{table*}[t]
\caption{Performance of the proposed feature extraction approach with three different classifier in comparison to different data augmentation strategies  by Equal Error Rate (EER$\downarrow$). Note that the state-of-the-art (SOTA) reference is from a fusion system. The reproduced system is a single system that uses the WORLD vocoder for feature extraction.}


\label{t:result}
\resizebox{\textwidth}{!}{%
\begin{tabular}{ccccclccc}
\hline
\multirow{2}{*}{Augmentation Category} & \multirow{2}{*}{Data Augmentation} & \multicolumn{3}{c}{Progress} &  & \multicolumn{3}{c}{Eval} \\ \cline{3-5} \cline{7-9} 
                                       &                                    & VAE     & OCSVM   & AnoGAN   &  & VAE    & OCSVM  & AnoGAN \\ \hline
-                                       & SOTA reference~\cite{dku}                     & 23.60   & -       & -        &  & 24.25  & -      & -      \\
-                                       & Reproduced~\cite{dku}                    & 23.07   & 23.07   & 23.04    &  & 24.73  & 24.21     & 24.20  \\
SpecAugment                                & FrequencyMasking                   & 23.09   & 23.04   & 23.04    &  & 24.77  & 24.21  & 24.20  \\
SpecAugment                                & TemporalMasking                    & 23.09   & 23.06   & 23.09    &  & 24.76  & 24.20  & 24.24  \\
WaveAugment               & AddNoise                           & 23.30   & 23.10   & 23.20    &  & 24.59  & 24.21  & 24.29  \\
WaveAugment               & AddReverb                          & 22.80   & 22.95   & 22.99    &  & 24.37  & 24.01  & 24.07  \\
WaveAugment                     & AdjustSpeed                        & 24.36   & 22.95   & 23.00    &  & 25.83  & 24.15  & 24.15  \\
WaveAugment                                      & De-emphasis                         & 22.92   & 23.11   & 23.09    &  & 24.43  & 24.20  & 24.21  \\
WaveAugment                                        & Pre-emphasis                        & 23.00   & 23.10   & 23.07    &  & 24.14  & 24.21  & 24.21  \\
   -                         & Audio Codec                        & \textbf{21.73}   & \textbf{21.53}   & \textbf{21.47}    &  & \textbf{23.40}  & \textbf{22.75}  & \textbf{22.71} \\ \hline
\end{tabular}%
}
\end{table*}

\begin{table*}[t]
\caption{Experiment results of different models under various bit-rate settings for audio codec, evaluated by Equal Error Rate (EER$\downarrow$).}
\label{t:opusresult}
\centering
\begin{tabular}{ccccclccc}
\hline
\multirow{2}{*}{Audio Codec} & \multirow{2}{*}{Bitrate} & \multicolumn{3}{c}{Progress} &  & \multicolumn{3}{c}{Eval} \\ \cline{3-5} \cline{7-9} 
                                       &                                    & VAE     & OCSVM   & AnoGAN   &  & VAE    & OCSVM  & AnoGAN \\ \hline

Opus                            & 16k                         & \textbf{21.73}   & \textbf{21.53}   & \textbf{21.47}    &  & 23.40  & \textbf{22.75}  & \textbf{22.71}  \\
Opus                            & 14k                        & 21.82   & 21.75   & 21.74    &  & \textbf{23.21}  & 22.91  & 22.90  \\
Opus                           & 12k                       & 22.21   & 21.77   & 21.68    &  & 23.66  & 22.95  & 22.89  \\
Opus                            & 10k                       & 22.02   & 21.87   & 21.81    &  & 23.59  & 23.13  & 23.08  \\
Opus                              & 8k                    & 23.67   & 23.11   & 23.10        &  & 25.59  & 24.99  & 24.97      \\ \hline
\end{tabular}%
\end{table*}

\section{Experiments}
\vspace{-3mm}
\label{sec:experiment}

\subsection{Dataset and Evaluation Metrics}
We conduct our experiments on ASVspoof 2019~\cite{asvspoof2019} and ASVspoof 2021~\cite{asvspoof2021} datasets. The classifiers were trained on the training set and the development set of the ASVspoof 2019 PA dataset, which are created through software simulations. The progress set and the evaluation set of the ASVspoof 2021 PA dataset,  recorded in real physical spaces, are used for evaluation. The progress set and the evaluation set are utterance-disjoint, and some recording setting factors are reserved exclusively for the evaluation set \cite{asvspoof2021}.

We employ the Equal Error Rate (EER) to evaluate the performance of the classifiers.
The output score of a classifier indicates the level of confidence in classifying the audio as a bonafide speech. To calculate the EER, a threshold is determined based on the output score, ensuring that the probability of missing a true positive is equal to the probability of false alarms. A lower EER indicates a better classification performance.

\subsection{Model and Feature Configurations}


The sampling rate of the data is 16 kHz. To extract replay channel response estimation features in~\cite{dku}, 
we set the FFT bins to 1024, the frame length to 50 ms and frame shift to 25ms. The first 512 dimensions were kept as features. The utterance level feature can be extracted by simply temporal averaging and PCA is applied to reduce the dimensions of features to preserve 98\% energy.

The replay channel response estimation follows the same preprocessing procedure as \cite{dku}: utilize the WORLD vocoder~\cite{morise2016world} to simulate the replay environment, then the subtraction result between original audio and simulated audio in log spectrogram level will be viewed as the feature. After temporal averaging, this feature will be fed into classifiers.

For the Opus codec, we set the sample rate to 16000Hz and the number of channels to 1, which corresponds to the audio file, for both the encoder and the decoder. And the application mode of encoder is chosen as 'VoIP', which gives best quality at a given bitrate for voice signals. We tried different bitrates from 16k to 8k, which are far more lower than the original audio, about 256k.

For data augmentation, different augmentation methods require different hyper-parameter settings. The maximum possible length of the mask of SpecAugment, including frequency masking and time masking, is $80$. The maximum proportion of time steps that can be masked is $100\%$. As for room scenario simulation, the SNR of AddNoise technique is $10$. Noise data and RIR data are extracted from the VOiCES dataset~\cite{voices} in our implementation. Coefficient of Pre-emphasis and De-emphasis is $0.97$. Speed Perturbation factors are set to $[0.9, 1.1]$. All data augmentation methods are implemented with TorchAudio\footnote{https://github.com/pytorch/audio}.





\subsection{Analysis of Data Augmentation}
We begin our analysis by examining the performance of spectrum augmentation. The results are presented in Table~\ref{t:result}. While the technique has been proven effective in the context of speech recognition, its use in the replay spoofing detection experiment did not result in any improvement. \textit{The experimental results suggest that neither time nor frequency masking is able to increase robustness to unseen replay artifacts.} However, it is worth noting that the small size of the dataset used in the experiment may have limited the effectiveness of spectrum augmentation, and further research with larger datasets could yield different results.

Moving on, we turn our attention to the performance of waveform augmentation. Table~\ref{t:result} presents the results of our assessment. Interestingly, we observed that adding reverberation and applying de-emphasis or pre-emphasis slightly decreased the EERs. It is possible that the added complexity introduced by these techniques helped the model better distinguish between genuine and spoofed audio. On the other hand, adding noise or adjusting speech speed appeared to be ineffective. These findings highlight the importance of careful selection and evaluation of augmentation techniques when working with audio data.




\subsection{Performance of Audio Compression/Codec}

\begin{figure}
    \centering
    \begin{subfigure}{0.5\linewidth}
        \centering
        \includegraphics[width=\linewidth]{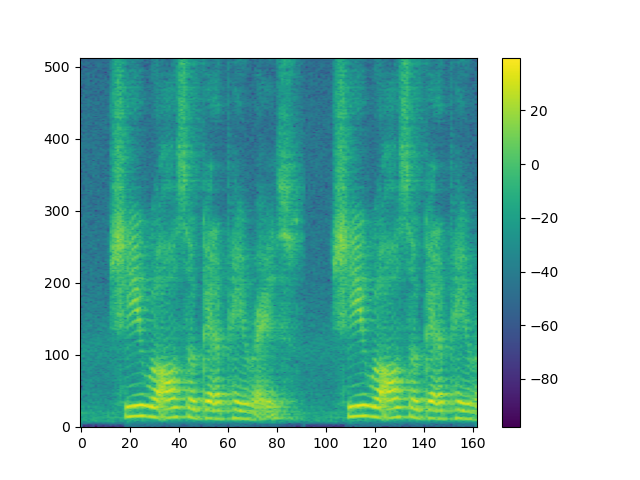}
        \caption{Original audio}
        \label{fig:sub1}
    \end{subfigure}%
    \begin{subfigure}{0.5\linewidth}
        \centering
        \includegraphics[width=\linewidth]{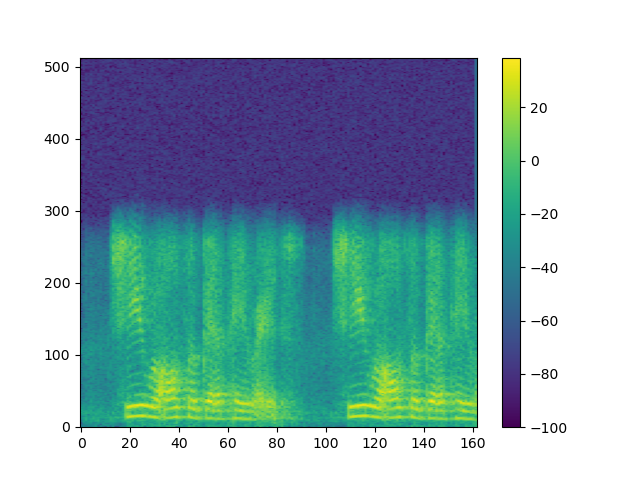}
        \caption{Opus compressed audio}
        \label{fig:sub2}
    \end{subfigure}
    \caption{Comparison of Mel spectrogram between the original audio and the 8k bitrates Opus compressed audio}
    \label{fig:fig2}
\end{figure}
In this subsection, we analyze the performance of audio compression-assisted feature in detail.

Firstly, we present the experimental results with varied bit-rates
in Table~\ref{t:opusresult}. It is worth noting that the results show that when the bitrate decreases from 16k to 10k, the EERs slightly increase for all 3 classifiers on both progress and eval datasets. This suggests that a lower bitrate can still preserve enough content and speaker information. However, when the bitrate is down to 8k, the EER deteriorates significantly. This is because when the bitrate is too low, it will impact audio quality and intelligibility. In fact, with 8k bitrates, the audio compressed by Opus actually loses almost all the high frequency information, as shown in Fig.~\ref{fig:fig2}. Consequently, high frequency noise cannot be extracted, which ultimately leads to poor results.

We can also argue that while a lower bitrate may still be able to preserve content and speaker information, it can have a significant impact on the audio quality and perceptibility. When the bitrate is down to 8k, the audio quality drops significantly, and this can have consequences for the overall result. Therefore, it is important to strike a balance between the bitrate and the audio quality, in order to achieve the best possible outcome.

In conclusion, while it is possible to achieve good results with lower bitrates, the quality of the compressed audio needs to be taken into account. A balance between bitrate and quality must be maintained to ensure optimal results.

\subsection{Performance of different classifiers}

Given the temporal averaged feature, we can employ a variety of classifiers, including VAE, OCSVM, and AnoGAN, to score the corresponding audio. The results of our tests show that OCSVM and AnoGAN outperform VAE, which was used in \cite{dku}, on the evaluation dataset. Specifically, the equal error rate (EER) achieved by OCSVM and AnoGAN are 24.21 and 24.20, respectively, which is $2.10\%$ and $2.14\%$ lower than that achieved by VAE on the same evaluation dataset relatively. Consequently, it is reasonable to conclude that both OCSVM and AnoGAN are superior to VAE in terms of evaluating the audio feature. However, it is worth mentioning that on the progress dataset, OCSVM and AnoGAN achieve nearly the same performance as VAE does. This finding suggests that OCSVM and AnoGAN have better generalization performance on a larger dataset. As a result, the application of these classifiers can be extended to a broader range of audio content, which can provide more reliable results.

The performance of the proposed feature extraction is presented in the last row of Table~\ref{t:result}. It achieves the lowest EERs regardless of the classifier used. In summary, the proposed feature extraction that employs audio compression for assistance can considerably improve detection performance. However, when designing the feature extraction using audio compression, an appropriate bitrate needs to be chosen.

\section{Conclusions}
This study proposes a feature extraction approach that utilizes audio compression for assistance. This is achieved by subtracting the reconstructed waveform from an audio codec from the original waveform. With the assistance of audio compression, we also explore the effect of different kinds of data augmentation techniques and compare their performance with three different one-class classifiers: variational autoencoder (VAE), one-class support vector machine (SVM), and AnoGAN, a deep convolutional generative adversarial network. The experiment conducted on the ASVspoof 2021 PA dataset suggests the effectiveness of the proposed approach, which achieves an equal error rate (EER) of 22.71\%. To the best of our knowledge, this is the lowest EER achieved on the dataset.

\bibliographystyle{IEEEbib}
\bibliography{refs}

\end{document}